\documentclass[onecollarge,natbib]{svjour2}
\bibpunct{[}{]}{;}{n}{}{,} 
\smartqed  
\usepackage{graphicx}
\usepackage{mathptmx}      
\def\bea{\begin{eqnarray}}
\def\eea{\end{eqnarray}}

\journalname{Few-Body Systems (EFB22)}
\begin{document}

\title{   Error Analysis of Nuclear Matrix Elements
~\thanks{
Supported by Spanish DGI (grant FIS2011-24149) and Junta de
Andaluc\'{\i}a (grant FQM225) and the Mexican CONACYT.  
}
}


\author{J. E. Amaro \and  R. Navarro P\'erez \and 
E. Ruiz Arriola    
}


\institute{J. E. Amaro \at
              Departamento
  de Fisica At\'omica, Molecular y Nuclear and Instituto Carlos I de
  Fisica Te\'orica y Computacional, Universidad de Granada, E-18071
  Granada, Spain              \email{amaro@ugr.es}          
\and  R. Navarro P\'erez \at
              Departamento
  de Fisica At\'omica, Molecular y Nuclear and Instituto Carlos I de
  Fisica Te\'orica y Computacional, Universidad de Granada, E-18071
  Granada, Spain              \email{rnavarrop@ugr.es}            
\and E. Ruiz Arriola \at
              Departamento
  de Fisica At\'omica, Molecular y Nuclear and Instituto Carlos I de
  Fisica Te\'orica y Computacional, Universidad de Granada, E-18071
  Granada, Spain              \email{earriola@ugr.es}           
} \date{Presented by J. E. A. at 22th European Conference On Few-Body
  Problems In Physics: EFB22 \\ 9 - 13 Sep 2013, Krakow (Poland)}

\maketitle

\begin{abstract}

We estimate the expected errors of nuclear matrix
elements coming from the uncertainty on the NN interaction.
We use a coarse grained (GR) interaction fitted to NN scattering data,
with several prescriptions for the long-part of the interaction, including 
one pion exchange and chiral two-pion exchange interactions.

\end{abstract}

\section{Introduction}
\label{intro}

We have recently made an error analysis of nuclear two body forces
based on a coarse graining of the unknown short range part of the NN
interaction that allows to quantify the uncertainties in the potential
parameters
\cite{Navarro12,Navarro12a,Navarro12b,Navarro13a,Navarro13b,Navarro13c,Navarro13d,Navarro13e,Navarro13f}.
Many nuclear structure calculations are carried out by diagonalization
of the many body nuclear Hamiltonian within the harmonic oscillator
shell model basis (possibly including the needed short range
correlations).  However, very little is known about the expected
accuracy of those calculations based on our lack of knowledge of the
input NN interaction.  In this talk we face the problem by deducting
and propagating two-body systematic and statistical errors to provide
a theoretical estimate of nuclear matrix elements and binding energy
uncertainties.  The impact of chiral Two Pion Exchange interactions
\cite{Epelbaum09,Machleidt11} in the evaluation of nuclear matrix
elements based on our error analyses can also be analyzed.  This may
help to set up {\it a priori} the needed accuracy to solve the many
body problem.

\section{Statistical and systematic errors of potential parameters}

Meaninful error estimates require to start with a NN potential fitting
the available data with $\chi^2/{\rm d.o.f.} \sim 1$. That potential
should be simple enough to allow the extraction of the errors in the
fitting parameters.  According to Aviles~\cite{Aviles72} one may
efficiently sample the unknown part of the interaction using a coarse
grained (GR) potential parameterized for each partial wave as a sum of
Dirac delta functions \cite{Navarro13b}.  Equivalently
\cite{Navarro13c,Navarro13e}, the potential can be written in operator
form as a linear combination of spin-isospin-angular operators as 
\begin{equation}
V(r)= \sum_{n=1}^{21}O_n\sum_{i=1}^N V_{i,n} \Delta r \delta(r-r_i) 
+ V_{L}(r)\theta(r-r_c) 
\end{equation}
extending the standard AV18 set \cite{Wiringa95}. Here $\Delta r=0.6
{\rm fm}$ is the resolution scale and $r_c=3{\rm fm}$ is the cut-off
radius. $V_L(r)$ is the long-range part of the interaction, including
one-pion exchage potential plus additional e.m. terms.  Our recent
partial wave analysis of NN scattering data below pion production
\cite{Navarro13c,Navarro13e} yielding $\chi^2/{\rm d.o.f.}  = 1.06$
with the GR potential, allowed to extract the statistical errors in
the fitted parameters. Alternatively one can fit directly to
phase-shift {\em pseudodata} for each partial wave.  Systematic errors
manifest when different high quality potentials show discrepancies in
the predicted phaseshifts.  The pseudodata are then obtained from the
computed phase-shifts for several high-quality potentials
\cite{Machleidt96,Stoks93,Wiringa95,Friar1993,Gross2008}.
The pseudodata are obtained as the average and standard deviation of
the computed phaseshift for a given energy.  A first conclusion can be
extracted: systematic errors are in general more than twice larger
than statistical ones in most of the observables.  The estimated error
in the nuclear binding energy per particle, obtained
 \cite{Navarro12} using several approaches for different nuclei in
a range of mass number, is $\Delta B /A = 0.1 - 0.4$ MeV.  In the next
section we show new results obtained for separated nuclear matrix
elements.

\begin{figure}
\begin{center}
\includegraphics[bb=140 440 560 660, width=15cm]{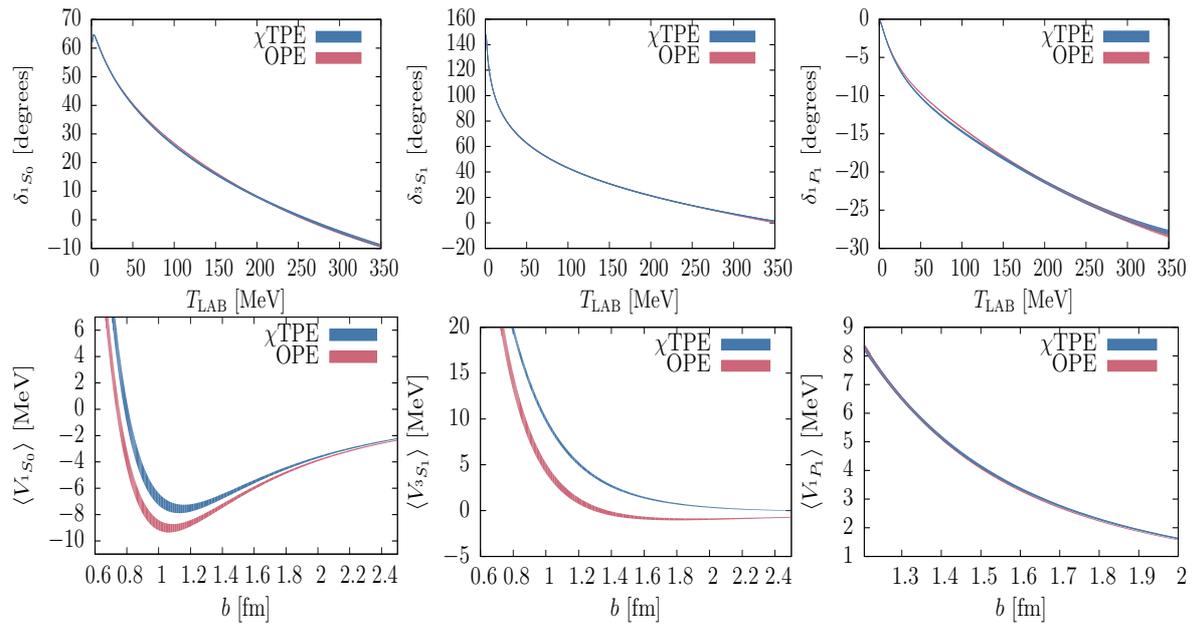}
\end{center}
\caption{Error bands of phaseshifts and potential expected values for 
the first partial waves $^1S_0, ^3S_1, ^1P_1$ 
obtained in two fits 
with  
$T_{\rm LAB} \le 350 {\rm MeV}$, $r_c|_{\rm OPE}=3 {\rm fm}$, 
$r_c|_{\rm TPE}=1.8 {\rm fm}$ (see text for details).}
\end{figure}

\begin{figure}
\begin{center}
\includegraphics[bb=140 440 560 660, width=15cm]{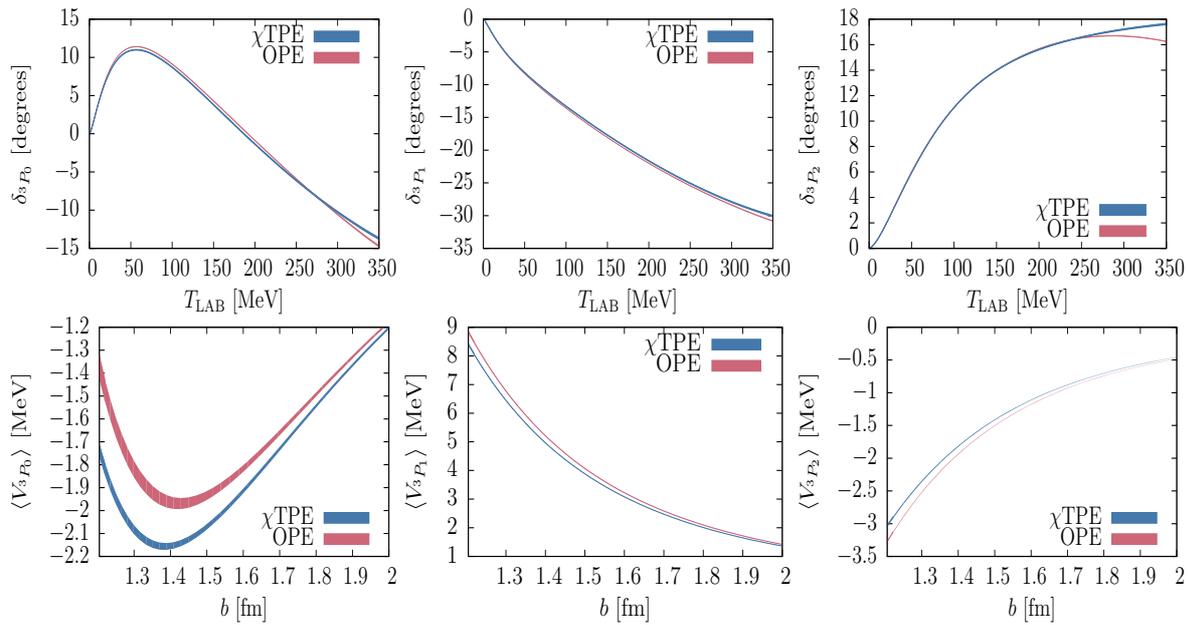}
\end{center}
\caption{The same as Fig. 1 for partial waves $^3P_0, ^3P_1, ^3P_2$
($T_{\rm LAB} \le 350 {\rm MeV}$,
$r_c|_{\rm OPE}=3 {\rm fm}$, 
$r_c|_{\rm TPE}=1.8 {\rm fm}$ ).}
\end{figure}

\section{Errors of nuclear matrix elements}

In Fig. 1 we show results for the first partial waves 
$^1S_0, ^3S_1, ^1P_1$ 
and in Fig. 2 for 
$^3P_0, ^3P_1, ^3P_2$. 
Statistical errors are
represented by an error band labeled OPE.  In
the upper panels we show the error of the phaseshifts as a function of
the energy. That error propagates to the expected values of the NN
potential energy on harmonic oscilator wave functions, displayed in the lower
panels as a function of the oscilator length.  The relative error in
the potential expected value is appreciably larger than in the phaseshift for
S-waves.

Modern chiral pertubation theory studies of nuclear structure
emphasize the universality of two-pion exchange ($\chi$TPE) for
intermediate to long distances.  Our model can easily be modified to
investigate the effect that the presence of $\chi$TPE in the NN
interaction would have on nuclear observables.  Therefore 
in Figs 1 and 2 we
show also results from a second fit with the potential modified to
include TPE in the intermediate region.  This involves to reduce the
cut radius to $r_c=1.8$ fm, and upward that
distance to define the potential as the sum of one- plus two-pion exchange.
$V_L(r)=V_{TPE}(r) + V_{OPE}(r)$.
Below $r_c$ the potential is again of the GR form.  This procedure
reduces the number of parameters in the fit, but increases the
$\chi^2/{\rm d.o.f.}$ value to 1.10 \cite{Navarro13f,Navarro13g}.  
As we can see from the
figures 1 and 2, the presence of $\chi$TPE in the potential produces
different results in the nuclear expected values for S-waves, taking
into account the statistical error.

\begin{figure}
\begin{center}
\includegraphics[bb=140 440 560 660, width=15cm]{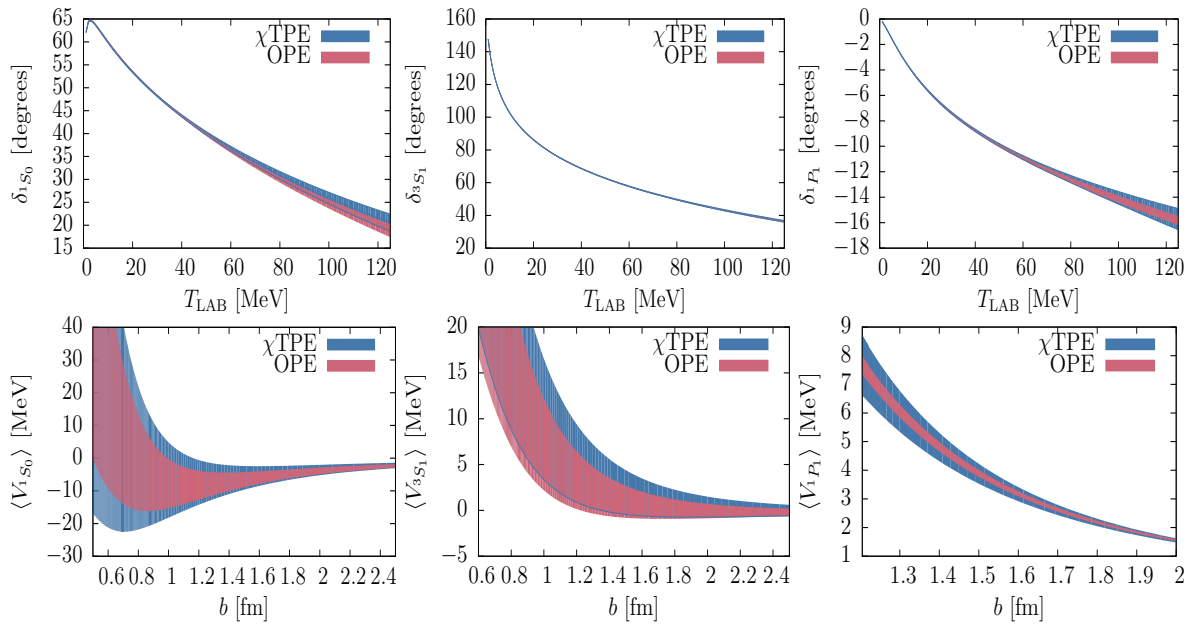}
\end{center}
\caption{
The same as Fig. 1 for 
$T_{\rm LAB} \le 125 {\rm MeV}$,
$r_c|_{\rm OPE}=1.8 {\rm fm}$, 
$r_c|_{\rm TPE}=1.8 {\rm fm}$ 
}
\end{figure}

\begin{figure}
\begin{center}
\includegraphics[bb=140 440 560 660, width=15cm]{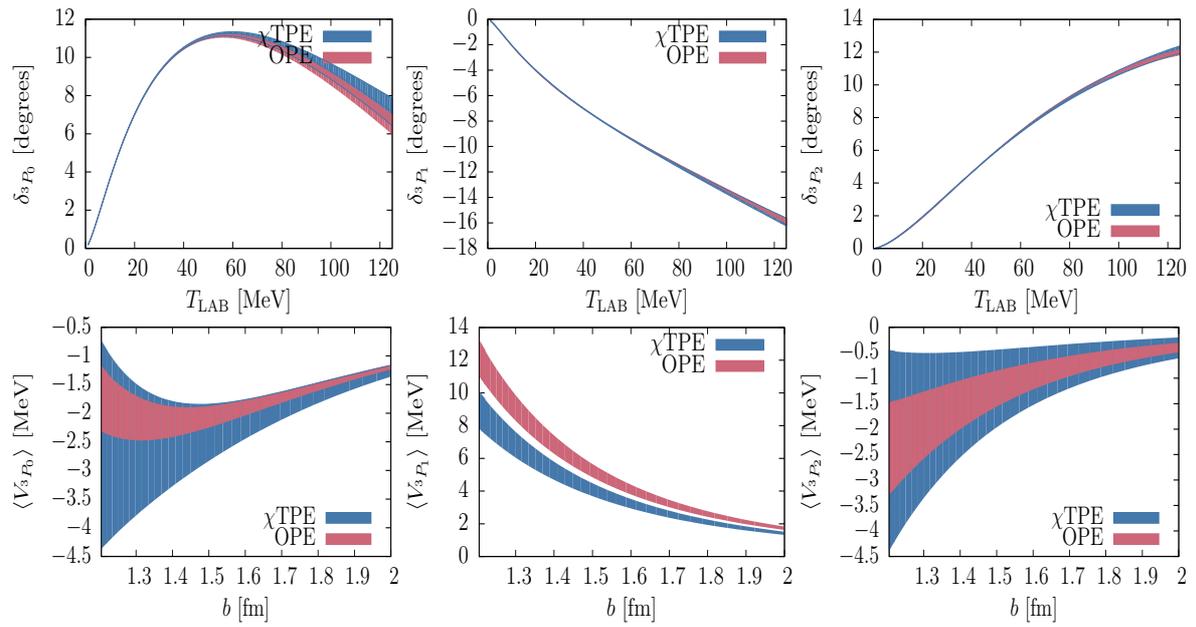}
\end{center}
\caption{The same as Fig. 2 for 
$T_{\rm LAB} \le 125 {\rm MeV}$,
$r_c|_{\rm OPE}=1.8 {\rm fm}$, 
$r_c|_{\rm TPE}=1.8 {\rm fm}$ 
}
\end{figure}

Recently, an optimized chiral potential has been fitted to np
scattering data by setting un upper cut-off in the LAB energy to
$E_{\rm LAB}= 125$MeV~\cite{Ekstrom13}. This corresponds to resolution
scale $\Delta r \sim 1.2 {\rm fm}$, or equivalently a low-momentum
interaction.  Shell model calculations~\cite{Coraggio09} with
low-momentum effective interactions are easier to work with than
G-matrix calculations, similarly to soft core NN
potentials~\cite{Afnan68}, and are able to provide an accurate
description of nuclear structure.  We explore the theoretical error
arising from that approach in Figs. 3 and 4.  Therein we show the
results from two fits of NN data for energy below 125 MeV, using a GR
potential with cut radius $r_c=1.8$ fm. In the first fit the
long-range interaction includes the OPE potential only. In the second
we add also $\chi$TPE. Errors are big in the expected value of the
potential, the larger being those obtained with the chiral potential.

\section{Conclusions}

Sumarizing, we have estimated  errors in nuclear matrix elements
coming from the uncertainty of the NN interaction. Errors 
are moderate, but not negligible, setting a theoretical 
limit to the precision that one can reach in nuclear physics calculations.
Low energy fits produce softer potentials but theoretical errors increase.


\end{document}